\documentclass[12pt]{article}
 \begin{document}
\pagenumbering{arabic}

 \medskip \begin{center} \Large {\bf $p$-adic Strings and their Applications\footnote{Based on
 a talk at the 2nd International Conference on $p$-adic Mathematical Physics, 15 - 21 September 2005, Belgrade, Serbia and Montenegro}}\\
  \normalsize                                 
 \bigskip \bigskip                   Peter G.O. Freund\footnote{freund@theory.uchicago.edu}\\
 \medskip \em{Enrico Fermi Institute and Department of Physics\\
 University of Chicago, Chicago, IL 60637}                                 
 							
 \bigskip \bigskip \bigskip
 
 \bf{Abstract}\\
 \end{center}
 
 The theory of $p$-adic strings is reviewed along with some of their
applications, foremost among them to the tachyon condensation problem in string
theory. Some open problems are discussed, in particular that of the superstring in 10 dimensions as the end-stage of the 26-dimensional closed bosonic string's tachyon condensation.

 \newpage

{\bf 1. Introduction}

\bigskip

Though, within errors, the values yielded by measurement of any observable can, of necessity, always be expressed as rational numbers, it is useful to treat the "true" spectra of many observables as continuous, i.e. endow them with a topology. This requires a completion of the field {\bf Q} of rational numbers. Normally, it is considered evident that this completion be Archimedean and one is thus led to the field ${\bf R}={\bf Q_\infty}$ of real numbers. But just as space, originally viewed as evidently commutative and Euclidean, turned out to offer interesting and even realistic new possibilities when these constraints were relaxed, so we can ask what would happen were we to entertain the possibility of a non-Archimedean ($p$-adic) \cite{K} completion of {\bf Q}.

As it turns out, the resulting theories are extremely simple and related to the Archimedean theory in a well-defined manner \cite{FO, FW, M, BFOW, BF}. One main advantage of this simplicity is that we can express the effective tachyon lagrangian in closed form, and therefore address directly the issue of tachyon condensation \cite{GS, S}. 

I will first review the main ideas and results of the theory of $p$-adic strings. Then I shall show how $p$-adic strings are used in understanding tachyon condensation. Surprisingly, some open issues in tachyon condensation are related to some long-open issues in $p$-adic string theory. I will therefore consider to what extent this relation can be used to further our understanding of both. 

\bigskip\bigskip\bigskip

{\bf 2. Review of $p$-adic and adelic strings}

\bigskip

To complete the field ${\bf Q}$ of rational numbers, we need Cauchy sequences and this requires a norm on ${\bf Q}$. Remarkably, all norms on ${\bf Q}$ are known. There is of course the familiar "absolute value" norm, which is Archimedean (i.e. for any pair of rational numbers $x$ and $y$, with $|x|<|y|$, there exists an integer $N$, such that $|Nx|>|y|$). For our purposes it will be convenient to denote this norm as $|x|_\infty$, rather than $|x|$. Besides this Archimedean norm, we have an infinity of non-Archimedean norms $|x|_p$ labeled by the prime numbers $p$. These $p$-adic norms measure the divisibility of the rational number $x$ by the prime $p$ (the more divisible, the smaller). According to  Ostrowski's famous theorem \cite{K}, the Archimedean norm $|x|_\infty$ along with the $p$-adic norms $|x|_p$ yield all the possible norms on the field ${\bf Q}$. 

In other words, there are infinitely many non-equivalent ways to complete the rationals ${\bf Q}$ to a field endowed with a topology. Of these completions precisely one is Archimedean and the infinite set of non-Archimedean completions are unambiguously labeled by the primes.

Using these $p$-adic norms to construct Cauchy sequences, we obtain locally compact completions ${\bf Q_p}$ of ${\bf Q}$. These are the  fields of $p$-adic numbers. Like ${\bf R} \equiv {\bf Q_\infty}$, all these ${\bf Q_p}$ are continua. Being locally compact, they have additive and multiplicative Haar measures \cite{GGP}, and as such we can integrate over these fields, just as we can over the reals. Moreover, these fields, just like the field of real numbers, have \cite{GGP} additive and multiplicative ${\bf C^*}$-valued characters. 

These observations allow us to construct all the $p$-adic string amplitudes. Indeed, the ordinary Archimedean Koba-Nielsen tree amplitudes are multiple integrals over ${\bf Q_\infty}$ of integrands built entirely of multiplicative characters of ${\bf Q_\infty}$. Replacing these characters of ${\bf Q_\infty}$ with characters of ${\bf Q_p}$, and replacing at the same time the integrals over ${\bf Q_\infty}$ by integrals over ${\bf Q_p}$, we obtain the tree amplitudes of a theory of $p$-adic strings. That these $p$-adic integrals do indeed represent the tree amplitudes of a string theory, is readily confirmed, by verifying that they possess the following features: they are a) meromorphic, b) M\"{o}bius-invariant, c) crossing symmetric, d) free of pairs of incompatible poles and e) they factorize. 

Unlike for the ordinary Archimedean string amplitudes, for these $p$-adic string amplitudes the integrations can be fully carried out \cite{BFOW, HS, FN}and the result expressed \cite{BFOW} in terms of elementary functions. 

The $p$-adic string, unlike its Archimedean counterpart, has no Hagedorn spectrum. Rather, its full spectrum is given by a single spin-zero tachyon. Therefore, without having to integrate out any higher modes, the $p$-adic amplitudes themselves can be used to generate the effective {\em target-space} tachyon action $S_p$ with the result,
\begin{equation}
S_p= \int d^dx {L}_p,  
\end{equation}
with the lagrangian $L_p$ given as,
\begin{equation}
{L}_p = \frac{p}{p-1}[-\frac{1}{2}\sigma p^{- \partial^2-1}\sigma + \frac{1}{g^2} \frac{p}{p+1} (1+\frac{g\sigma}{p})^{p+1} -\frac{\sigma}{g} - \frac{1}{g^2} \frac{p}{p+1}]+ C_p. 
\end{equation}
Here $\sigma$ is the real scalar tachyon field propagating in $d$-dimensional Archimedean space-time (the familiar value $d=26$ of the critical dimension of Archimedean bosonic strings can be shown, using adelic methods, to be needed in the $p$-adic case as well) and $g$ is the ($p$-dependent) coupling constant. Notice that the kinetic term is exponential in derivatives, making this lagrangian non-local. This non-locality has the consequence that along with the tachyon pole at $m^2 = -2$, the $p$-adic string amplitudes also have equally spaced {\it complex} poles at $m^2 = -2 + i 4\pi N/ \ln p$ for all integer $N$. The, in a lagrangian irrelevant, additive constant $C_p$ has been introduced here for future convenience. It is important to emphasize again that this lagrangian gets integrated over the full {\em real} $d$-dimensional target space. It remembers its $p$-adic origins through its non-locality and through the explicit appearance of the prime $p$ in the lagrangian. Momenta and target-space coordinates remain real (or complex upon suitable continuation), in other words Archimedean. Only the worldsheet is $p$-adic.  

The field-redefinition, 
\begin{equation}
\phi= 1 +g\sigma/p
\end{equation}
allows us to rewrite this lagrangian as
\begin{equation}
{L}_p = \frac{1}{g^2} \frac{p^2}{p-1}[-\frac{1}{2}\phi p^{- \partial^2/2}\phi + \frac{1}{p+1}\phi^{p+1}].
\end{equation}

Notice that this $L_p$ vanishes for $\phi=0$. This has been achieved by setting the arbitrary constant 
\begin{equation}
C_p=-\frac{1}{2g^2}\frac{p^2}{p+1},
\end{equation}
This will make the dicussion of tachyon condensation particularly transparent.

Exact solitonic solutions of the field equation
\begin{equation}
p^{- \partial^2/2} \phi =\phi^p
\end{equation}
can be found. These solutions are important in understanding tachyon condensation, as will be shown in the next section.

What is the physical meaning of the $p$-adic strings? For an open string, the worldsheet's boundary is no longer the real line as in the ordinary Archimedean case, but is now postulated to be the $p$-adic line. What kind of worldsheet has such a boundary? The answer to this question is a certain $p$-adic symmetric space \cite{Z}, a Bruhat-Tits tree. It is as if we were discretizing the worldsheet, for after all a Bruhat-Tits tree is a Bethe lattice. There is here an intrinsic difference between the target space and the worldsheet. We will return to this problem in the following sections.

Finally, before concluding our very brief review of $p$-adic strings, let us notice that any $p$-adic string requires the choice of the prime $p$. Now, even if the mentioned non-locality of the theory were not a major problem, were we trying to use $p$-adic strings for phenomenology, we would be in the preposterous position of having to ask our experimental colleagues to {\it measure} the prime number $p$ which underlies the theory. It would be much more satisfactory if somehow we could treat all primes on an equal footing, and this includes the prime "at infinity" which corresponds to the usual Archimedean strings. This can be achieved by considering the so-called {\it adelic} strings \cite{FW, M, BF}. Adelic strings establish such a relation between all $p$-adic strings and the Archimedean strings. Specifically, by taking the (suitably regularized) product of the 4-point amplitudes of {\it all} $p$-adic strings and then multiplying the result with the ordinary Veneziano amplitude, we obtain a constant. In the limit in which all Archimedean and non-Archimedean string couplings are set equal to one, the value of this constant itself turns out to be equal to one. Strictly speaking this result holds at the tree level, but its generalization to the full amplitudes has also been proposed \cite{BF}. 

\bigskip\bigskip\bigskip

{\bf 3. Tachyon Condensation}

\bigskip

A tachyon in the spectrum of a quantum system is a signal that we are expanding it around a false vacuum. It is therefore important to understand how the system finds its true vacuum. The bosonic string --- whether Archimedean or $p$-adic --- has a tachyon in its spectrum and as such we know that it is expanded around a false vacuum. This expansion corresponds to small $\sigma$ field, and indeed $\sigma=0$, or equivalently $\phi=1$ is a solution of the field equation (6). Being constant over all of target space, it corresponds to a target space volume-filling D-brane, specifically a D-(d-1)-brane. As expected this D-brane is the one around which we quantized the string and found the tachyon. The energy density of this D-brane can be read off Equation (4),
\begin{equation}
   T^p_{d-1}= -L_p(\phi=1)= \frac{1}{2g^2}\frac{p^2}{p+1}.
\end{equation}

The linearized field equation around $\phi=1$ has plane wave solutions with with momentum constrained to the open string tachyon mass-shell.

The configuration $\phi=0$ is also a solution of the field equation (6). With the choice (5) for $C_p$, we had $L_p(\phi=0)=0$, so the energy of this solution vanishes. There is no D-brane in this case, and there are therefore no plane wave solutions with finite momentum squared, in other words no open-string-like perturbative excitations around this $\phi=0$ vacuum. All that is left are the more fundamental closed string excitations, but these can not be readily found in this not yet second-quantized picture. 

Finally \cite{BFOW}, the field equation (6) admits interesting soliton solutions. Specifically, the configuration, 
\begin{equation}
 \phi(x)=\Pi _{j=q+1} ^{j=d-1} f(x^j) := F^{d-q-1}(x_\perp),
\end{equation}
with
\begin{equation}
 f(\eta)= p^{\frac{1}{2(p-1)}} exp (-\frac{1}{2}\frac{p-1}{p\ln p}\eta^2), ~~~~ 
 x_\perp= (x^{q+1}, ... ,x^{d-1}),
\end{equation}   
solves the field equation (6), and its energy is localized around the hyperplane $x_\perp=0$.
Call $T_q$ the energy per unit $q$-volume of this solution. It is finite, exactly calculable, and has the remarkable property \cite{GS, S} that the ratio $T_q/T_{q-1}$ is independent of $q$. This is precisely what happens for the ratio of the tensions of an ordinary D-$q$-brane and an ordinary D-$(q-1)$-brane in Archimedean bosonic string theory. Moreover these ordinary D-branes are also of solitonic nature. It then stands to reason to interpret \cite{GS, S}
the solitons of Eqs. (8), (9) as the D-$q$-branes of the $p$-adic string. In support of this interpretation, one can show \cite{GS, S} that the spectrum and dynamics of fluctuations around these D-$q$-branes coincides with what one expects for the open string quantized on the D-$q$-brane.  

This interpretation is indirect in the following sense. The whole argument was made at the level of the effective tachyon action. This involves the theory of a {\em real} scalar field in ordinary Archimedean space-time, with the $p$-adic features encoded, as explained above, through the non-locality of the lagrangian, and the explicit appearance of the prime $p$ in it. The fluctuations around the D-brane are identified with $p$-adic strings also at the tachyon lagrangian level. This description lacks the intuitive picture of a string moving through space-time with its ends attached to the D-brane. This is just as good, since in the $p$-adic case, the world lines of the string's ends are $p$-adic lines and as such could not be fit into a D-brane all of whose coordinates are valued in the {\em Archimedean} field of real numbers.

The just described features of the effective tachyon action which were readily obtained in the $p$-adic string case, correspond to some very general conjectures about tachyon condensation \cite{S}. The advantage in studying $p$-adic strings is precisely that one can easily do analytic calculations and verify these conjectures.

It is interesting to note that in the (local) limit $p\rightarrow 1$ the p-adic string's effective action has been found \cite{GERS}, \cite{G} to approximate up to terms with two derivatives, the effective action of the Archimedean bosonic string calculated using boundary string field theory.
\bigskip

In the usual {\em Archimedean} bosonic string, when $N$ D-branes coincide, a $U(N)$ symmetry with the attendant Chan-Paton (CP) rules is known to emerge. Does something similar happen in the $p$-adic case as well? A possible theory of $p$-adic strings with CP rules has been proposed \cite{BF}. It had the undesirable feature that for $N=1$ its amplitudes did not reduce to the those of the theory without CP rules described above. This problem can be corrected by constructing an alternative $p$-adic string with CP rules as follows. In the lagrangian (4) let $\phi$ denote not a unique real scalar field, but an $N \times N$ matrix of such fields, with $L_p$ obtained from Eq. (4) by taking the matrix trace. Then the theory clearly reduces to what it should for $N=1$. The question that remains to be answered is: what is the worldsheet description of the $p$-adic theory whose effective tachyon action is given by this matrix-alternative construction?
\bigskip

In the Archimedean case, unstable D-branes decay into closed strings, so that the open string theory contains the whole dynamics of D-branes and even much of the closed string dynamics. This is the content of Sen's far-reaching {\em open string completeness conjecture}. Closed bosonic strings also have a tachyon, but here the tachyon condensation is a much more complex issue. One might at first think that also for closed strings the $p$-adic case will allow for an exact treatment. Unfortunately this is not the case. The main difference between the open and closed bosonic Archimedean strings is that to obtain the tree amplitudes one integrates over the field ${\bf R}= {\bf Q_\infty}$ for the open strings and over the field {\bf C} of complex numbers for the closed strings. The field {\bf C} being a quadratic extension of the field {\bf R}, at the $p$-adic level one can try to use a quadratic extension of the field ${\bf Q_p}$. But the Archimedean and non-Archimedean situations are different. The field {\bf C} of complex numbers is algebraically closed, whereas quadratic extensions of the field ${\bf Q_p}$ of $p$-adic numbers -- just like the quadratic extensions of the "global" field {\bf Q} of rational numbers --- are neither unique nor algebraically closed. 

If one uses these quadratic extensions of ${\bf Q_p}$ anyway, one obtains sets of consistent $n$-point amplitudes, with the corresponding 4-point amplitudes in just as nice an adelic product relationship with the Archimedean {\em closed} string amplitudes, as was the case for ${\bf Q_p}$ and the open string 4-point amplitudes. This may lead us to think that these are the proper closed $p$-adic strings. But this is not necessarily so. First of all we have to ask what would have happened, had we taken not a quadratic, but a higher algebraic extension of the field ${\bf Q_p}$. These would be adelically related not to open or closed string Archimedean 4-point amplitudes, but rather to products of various powers of open and closed string amplitudes, the precise values of these powers depending on precisely which of the infinitely many possible algebraic extensions is being considered. A product of $n_r$ open and $n_i$ closed string amplitudes, needed in such an adelic product does not seem to make any physical sense when $n_r>1$, or $n_i>1$, or $n_r n_i=1$. It appears that we are dealing in all these cases with ever more exotic forms of open string amplitudes. This is most easily confirmed by noticing that for all these cases one can construct an effective tachyon action, with the corresponding field equations admitting various solitonic D-brane solutions, a clear sign that one is dealing with --- admittedly more exotic kinds of --- open strings. To get to what are "true" closed strings, one may have to switch from the boundaries of Bruhat-Tits trees to the boundaries of infinite graphs that are no longer trees, but contain loops as well \cite{CMZ}. Unfortunately effective tachyon actions are no longer available in this case.
\bigskip

There remains the question as to the nature of the closed bosonic string's vacuum. If anything like what happened for the open bosonic string, happens here as well, then in this vacuum there should be no closed {\em bosonic} strings. In other words, there should be no 26-dimensional graviton! Then all of 26-dimensional space should disappear as well. What is {\em it} replaced by? Obviously by a system without a tachyon. But to eliminate tachyons we must end up with something closely resembling supersymmetry, as has been shown some time ago by Kutasov and Seiberg \cite{KZ}. The most natural outcome would then be the 10-dimensional space of the superstring, with its graviton. This is in line with an old idea of the superstring as a vacuum of the bosonic string \cite{F2, E, R}. This way, not only would the five types of 10-dimensional superstring theories be unified, but the odd-man-out, the 26-dimensional bosonic string would be unified with its supersymmetric partners as well. We could start with bosons only, and then fermions would be forced upon us, so as to get rid of the tachyons. 
\bigskip

As in the Archimedean case, in the $p$-adic case as well, an antisymmetric rank-2 tensor field $B$ can be included \cite{GK},with a resultant non-commutative target space geometry underlying the effective tachyon field action. This takes some doing and involves $p$-adic sign-functions in a manner similar to that used in \cite{BFOW} when introducing CP rules. There, as was already mentioned, for $N=1$ one did not recover the ordinary theory without CP rules. Here a similar problem arises, for if one sets $B=0$ one does not recover the ordinary string amplitudes without $B$ field. It would be interesting to see whether there is a connection between these singular behaviors observed when introducing CP rules and non-vanishing $B$ field.
\bigskip

Finally, although not directly related to the issue of tachyon condensation, I should mention one more, potentially important, adelic feature of string theory. In the compactifications of various string theories one encounters \cite{N, SCH} scalar fields which live on the so-called Narain moduli spaces (similar spaces were already encountered \cite{CJ} in toroidally compactified supergravities). All these Narain moduli spaces are double coset spaces of the form $G({\bf Z})\backslash G({\bf R})/ K_{\bf {R}}$, with $G({\bf R})$ a non-compact real form of some Lie group, $K_{\bf R}$ its maximal compact subgroup, and $G({\bf Z})$, which ultimately functions as duality group, the suitably defined discrete group $G$ over the integers. Typical cases are $G=O(22,6; {\bf R})$ for heterotic string compactifications and $E_{7,7}$ for certain type II compactifications. The remarkable feature is that these double coset spaces can, on the basis of the strong approximation theorem \cite{GGP}, also be represented  as the {\em adelic} double coset spaces $G({\bf Q})\backslash G({\bf A})/K_{\bf A}$, with $G({\bf A})$ the group $G$ over the adeles, $K_{\bf A}$ {\em its} maximal compact subgroup, and $G({\bf Q})$, the group $G$ over the rationals. (this equivalence was used in the theory of scattering on the hyperbolic plane \cite{F}). This observation suggests that a proper understanding of string dualities should involve an adelic formulation. At the same time it should be mentioned, that at the $p$-adic level string compactifications have not yet been explored.

\bigskip\bigskip\bigskip

{\bf 4. Outlook}

\bigskip

Originally, the possibility of the $p$-adic and adelic strings reviewed above, was recognized, once the importance of algebraic geometry in string theory became clear. After all, the worldsheets of closed strings are compact Riemann surfaces, all of which are algebraic curves (one {\em complex} dimension), as was known already to Riemann and then rigorously proved by Weyl. On the other hand, algebraic geometry is intimately tied in with number theory, as had been recognized by mathematicians for over a century. This immediately suggested the generalization of strings to the $p$-adic and adelic case. The most ambitious possibility is that in nature we are really dealing with adelic strings, and Archimedean strings are only an approximation to these adelic strings, say at low energies. This far-reaching point of view, though still tenable, has not led to any major progress. Part of the reason for this is that $p$-adic strings in their current form do not seem to take well to supersymmetry. The other difficulty is that adelic product formulas have only been obtained for 4-point tree amplitudes. Though ways of going beyond tree level have been suggested \cite{BF}, they are more complicated and have not led to any major insights.

Then came the realization \cite{GS} that the remarkable intuition \cite{S}about tachyon condensation can be easily checked in the $p$-adic case. In this context non-Archimedean strings have regained some popularity, but essentially as "toy models." 

If we insist on a fundamental role for $p$-adic and adelic strings, then we first have to master the supersymmetric case and higher orders in perturbation theory. 

There is also the issue whether it is consistent to go $p$-adic only on the worldsheet, while leaving the coordinates of target space (and therefore their conjugate momenta) Archimedean. We have seen how the effective tachyon actions of $p$-adic strings handle this issue in a rather indirect and non-intuitive manner. But could one entertain the possibility of a world, which is $p$-adic both on the worldsheet and in target space? This possibility has been advocated \cite{V, ADFV}, but within its context even the strings, where everything started, have not yielded a meaningful theory. 

I am aware of the fact that what I have just written has a pessimistic tinge to it. To end on a positive note, let me stress that the adelic string product formulas and the solitonic D-brane solutions of the field equations derived by extremizing the effective tachyon action, are both mathematically beautiful and physically interesting. As far as Physics is concerned, they have contributed a useful laboratory where one can test tachyon condensation ideas. I simply cannot think of all this as a sequence of amusing accidents and dismiss it as a "toy model." I think more work on this is both needed and worthwhile. After all, to put things in perspective, the main {\em kmown} virtues of the whole edifice of {\em Archimedean} string theory are {\em its} mathematical richness and {\em its} role as a useful laboratory where one can test ideas about black holes. 
\bigskip

{\bf Acknowledgements} 
\medskip

I wish to thank Prof. Branko Dragovich for inviting me to participate in the organization of and to speak at the Belgrade Conference on $p$-adic Mathematical Physics.

 \end{document}